\documentclass[aps,prb,twocolumn,superscriptaddress,floatfix,nofootinbib,longbibliography]{revtex4-2}

\usepackage{graphicx}
\graphicspath{{figures/}}
\usepackage{amsmath}
\usepackage{amssymb}
\usepackage{bm}
\usepackage{xcolor}
\usepackage[colorlinks=true,linkcolor=blue,citecolor=blue,urlcolor=blue]{hyperref}

\newcommand{\braket}[2]{\langle #1 \vert #2 \rangle}
\newcommand{\ketbra}[2]{\lvert #1 \rangle\!\langle #2 \rvert}

\newcommand{\rket}[1]{\lvert #1 \rangle_{\mathrm{R}}}
\newcommand{\lket}[1]{\lvert #1 \rangle_{\mathrm{L}}}
\newcommand{\lbra}[1]{\langle #1 \rvert_{\mathrm{L}}}

\newcommand{\nbel}{n_{<}}                     
\newcommand{\Eg}{E_{g}}                       
\newcommand{\half}{\tfrac{1}{2}}
\newcommand{\third}{\tfrac{1}{3}}

\newcommand{\creo}[1]{c^{\dagger}_{#1}}
\newcommand{\anno}[1]{c_{#1}}

\newcommand{\gpes}{\gamma_{\mathrm{PES}}}     
\newcommand{\gdem}{\gamma_{\mathrm{dem}}}     

\newcommand{\rdm}{\rho}                        

\begin{document}

\title{Fate of the non-Abelian Moore--Read manifold under the non-Hermitian skin effect}

\author{Jiaxuan Guo}
\email{guojx@stanford.edu}
\affiliation{Department of Applied Physics, Stanford University, Stanford, California 94305, USA}

\author{Simin Nie}
\affiliation{Department of Materials Science and Engineering, Stanford University, Stanford, California 94305, USA}

\author{Xiting Zhang}
\affiliation{Department of Materials Science and Engineering, Stanford University, Stanford, California 94305, USA}

\author{Fritz B. Prinz}
\email{fprinz@stanford.edu}
\affiliation{Department of Materials Science and Engineering, Stanford University, Stanford, California 94305, USA}
\affiliation{Department of Mechanical Engineering, Stanford University, Stanford, California 94305, USA}

\date{\today}

\begin{abstract}
We study a non-Abelian Moore--Read fractional Chern insulator under a translation-preserving,
nonreciprocal deformation that generates the non-Hermitian skin effect under open boundaries. The
model combines the imaginary-gauge Hatano--Nelson deformation with a kagome-lattice three-body
interaction designed to stabilize Moore--Read order at $\nu=\half$. Our primary diagnostic is the
biorthogonal $(2,4)$-admissible particle-entanglement counting of the sixfold Moore--Read
manifold, the standard Moore--Read fingerprint. Across three sizes
($N=16,20,24$), the counting locks to the clean references $1308$, $3965$, and $9282$ over finite
nonreciprocity windows through $\gamma\le0.55$, $0.65$, and $0.74$, respectively,
with positive reference-rank entanglement gaps. Within every reported window the count is unchanged
by the spectral readings tested; at $N=16$ it is also unchanged across three reduced density
operators, with all $15$ combinations returning $1308$. The sixfold pattern for even $N_f$
and the adiabatically tracked Ising-odd doublet remain separated over the tested range
$\gamma\le0.6$. Beyond a geometry-dependent threshold the
instantaneous-lowest-six reference-rank gap drops
sharply and its counting destabilizes. At $N=24$ a sector-$0$ branch pair becomes complex conjugate
over a narrow interval inside the delocking bracket; both continuations through the interval are
delocked at the tested PES points $\gamma=0.76$, $0.77$, and $0.80$.
A same-lattice Abelian $\nu=\third$ Laughlin realization retains its counting to $\gamma=1.0$, so its
counting is the more robust. On the torus the eigenstates remain
extended; under open boundaries the right and left states skin-localize at opposite edges while the
biorthogonal particle-entanglement spectrum is invariant under the imaginary-gauge similarity, so the
torus and the open cylinder probe the same deformation under periodic and open boundaries.

\end{abstract}

\maketitle

\section{Introduction}
\label{sec:intro}

A fractional Chern insulator (FCI) is the lattice realization of the fractional quantum Hall
effect: interacting electrons in a nearly flat, topologically nontrivial band organize into a
fractionalized topological state without an external magnetic field~\cite{tang2011,sun2011,neupert2011,sheng2011,regnault2011,bergholtz2013,parameswaran2013}.
Most fractional states are Abelian: braiding two quasiparticles returns the system to the
same state up to a phase, as in the Laughlin series~\cite{laughlin1983}. The Moore--Read
$\nu=\half$ Pfaffian state is the canonical exception~\cite{moore1991,read1999beyond}: its
quasiholes are non-Abelian, carrying Majorana zero modes, and braiding them rotates the system
within a degenerate manifold rather than multiplying it by a phase. On a torus this non-Abelian
order appears as a sixfold ground-state degeneracy and, more sharply, as a
counting rule in the particle-entanglement spectrum (PES)~\cite{li2008,sterdyniak2011}, the
diagnostic that distinguishes a genuine Moore--Read state from Abelian or symmetry-broken
competitors that can mimic its degeneracy.

In parallel, non-Hermitian (NH) physics describes open and driven
quantum matter~\cite{ashida2020,bergholtz2021}. One lattice signature is the
non-Hermitian skin effect (NHSE): with nonreciprocal hopping, extensively many eigenstates
pile up at a boundary and the spectrum migrates into the complex plane~\cite{hatano1996,yao2018,okuma2020}.
In interacting fermionic systems the Pauli principle modifies the skin effect, which still leaves a
density imbalance and entanglement signatures~\cite{mu2020manybody,alsallom2022,kawabata2023skin}.
Because left and right eigenvectors no longer coincide, we use biorthogonal observables unless
explicitly stated otherwise~\cite{kawabata2019,bergholtz2021}. We ask how the skin effect changes a
fractionalized, topologically ordered ground state.

Non-Hermitian Hamiltonian studies of fractional quantum Hall and Chern-insulator states have mainly
treated Abelian Laughlin order~\cite{yoshida2019,bergholtz2025skin}. Special non-Hermitian
pseudopotential constructions have also connected parameterized ground states to Laughlin and
Moore--Read limits~\cite{hasebe2009}, but do not address a skin-generating deformation or the biorthogonal Moore--Read
PES.
Hermitian Moore--Read FCIs range from early flat-band models~\cite{liu2013nonabelian,wang2015nonabelian}
to recent moir\'e and cold-atom realizations~\cite{nonabelian_moire2024,pfaffian_bosons2026}, with
related band-geometry and solvable constructions~\cite{kagome_nonabelian2025,nonabelian_solvable2026}.
To our knowledge, this is the first explicit study of a non-Abelian Moore--Read FCI manifold,
diagnosed through a biorthogonal Moore--Read PES, under a translation-preserving nonreciprocal
deformation that generates the skin effect.

Here ``non-Abelian'' means Moore--Read topological order, not the non-Abelian band braiding or gauge
structure used in single-particle non-Hermitian topology~\cite{wang2021braiding,hu2021knots}.

Our construction separates the interaction from the non-Hermiticity. The
interaction is purely real and three-body ($U_2=0$, $U_3=1$), a lattice analogue of the Moore--Read
three-body parent pseudopotential that stabilizes Moore--Read order in the kagome flat band. The
non-Hermiticity is a uniform nonreciprocal
hopping equivalent to a complex shift $k_1\to k_1+i\gamma$ of the dimensionless momentum
component $k_1=k\cdot a_1$ of the kagome Bloch Hamiltonian,
projected onto its lowest band. Because the hopping preserves
lattice translation, momentum stays a good quantum number and the Hamiltonian remains block
diagonal in momentum sectors. We diagonalize the biorthogonal complex spectrum, build the
six-state density operator from left/right eigenpairs, and compute the non-Abelian $(2,4)$-admissible PES
counting as a function of the nonreciprocity $\gamma$.

\begin{figure}[!t]
  \centering
  \includegraphics[width=\linewidth]{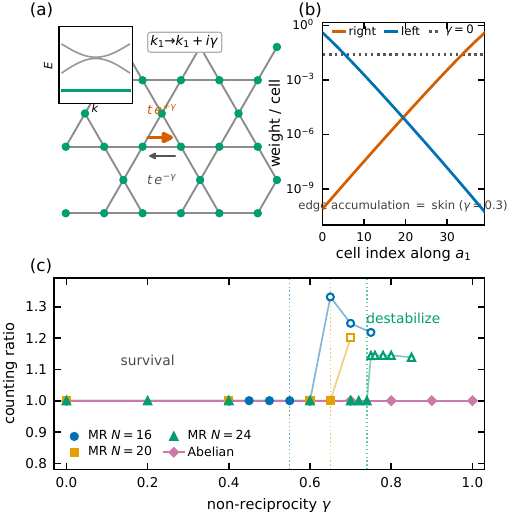}
  \caption{Model and results. (a) Kagome flat band with the imaginary-gauge (Hatano--Nelson) deformation
    $k_1\to k_1+i\gamma$, the generator of the non-Hermitian skin effect. The inset is a schematic
    of the flattened band structure, with the flat lowest band (green) isolated from the two
    dispersive bands. (b) The resulting skin effect on the underlying single-particle carrier. With an open
    $a_1$ boundary ($\gamma=0.3$), the right eigenstates pile at one edge and
    the left eigenstates at the other (mean weight per cell over all states, log scale), compared with the
    flat $\gamma=0$ profile; the full set of single-particle skin signatures is in the Supplemental
    Material, Sec.~S5. (c) The
    Moore--Read $(2,4)$ entanglement-counting ratio (measured over the clean reference) stays
    locked at $1$ across $N=16,20,24$ over a finite, size-dependent nonreciprocity window (dotted
    lines mark the window edges). The main-convention count changes past the distinct threshold
    $\gpes$ (shown for the instantaneous lowest-six reading; Sec.~\ref{sec:coda}), while a same-lattice Abelian
    $\nu=\third$ Laughlin state ($N=24$) stays locked throughout.}
  \label{fig:overview}
\end{figure}

Over a finite window, the non-Abelian $(2,4)$ PES counting locks index-by-index to the clean
Moore--Read reference at each of $N=16,20,24$, independently of the tested spectral reading and
reduced density operator (Sec.~\ref{sec:survival}). The separately defined $\gpes$ brackets locate
delocking under the main convention and need not share a window endpoint. The tracked Ising-odd doublet
gives an independent energy-level check, and on an open boundary the interacting state develops the
imaginary-gauge skin (Sec.~\ref{sec:baseline} and Supplemental Material Sec.~S5). Beyond $\gpes$ the
instantaneous-lowest-six entanglement gap drops sharply and its counting leaves the reference. At
$N=24$ this PES delocking precedes the later
reordering of $\mathrm{Re}\,E$ between momentum sectors. The continued and instantaneous lowest-six
subspaces are distinguished in Sec.~\ref{sec:coda}. Using the same lattice, non-Hermiticity, and code, we find that an
Abelian $\nu=\third$ Laughlin state instead keeps its $(1,3)$ counting locked to the largest
$\gamma$ scanned.

The $(2,4)$ counting and the Ising even-odd degeneracy provide the evidence for non-Abelian order. The counting integer locks across the window.

\section{Model and diagnostics}
\label{sec:model}

\subsection{Flat-band Moore--Read interaction}

We use the standard tight-binding Hamiltonian on a kagome lattice with three sites per unit cell. It
combines real nearest-neighbor hopping with equal-strength imaginary chiral nearest-neighbor hopping
that opens a topological gap. Its lowest band can be made nearly flat, with a large gap to the two
upper bands~\cite{sun2011,tang2011} (explicit form and parameters in
Appendix~\ref{app:explicit}). An $L_1\times L_2$ torus carries $3L_1L_2$ single-particle
orbitals across the three kagome bands. We keep the lowest band and project the interaction onto
it. Throughout, $N\equiv L_1L_2$ denotes the number of lowest-band orbitals retained after
projection ($N=16,20,24$ for the $4\times4$, $5\times4$, $6\times4$ tori), and the
$(2,4)$-admissibility counts are evaluated on these $N$ orbitals. The flattening replaces
the band dispersion by a constant, so the lowest band is exactly degenerate and only the projected
interaction determines the physics~\cite{regnault2011,wu2013bloch}.

The interaction is a pure three-body contact term,
\begin{equation}
  \begin{gathered}
    H_{\mathrm{int}} = U_3 \!\!\sum_{i<j<k}\!\! P_3(i,j,k) \; + \; U_2\!\!\sum_{i<j}\!\! P_2(i,j), \\
    U_3=1,\quad U_2=0 ,
  \end{gathered}
  \label{eq:Hint}
\end{equation}
where $P_3$ projects three particles onto the shortest-range relative configuration and $P_2$
is the corresponding two-body projector. The pure three-body term ($U_2=0$) is a lattice analogue of
the Moore--Read parent pseudopotential~\cite{moore1991,read1999beyond}. It favors the sixfold
Moore--Read subspace, near-degenerate at these finite sizes (Sec.~\ref{sec:baseline}), and gaps the rest. At filling $\nu=\half$ the
fermion number is $N_f=N/2 = 8,10,12$ on the $4\times4$, $5\times4$, $6\times4$ tori; these are
even, which gives the sixfold Pfaffian manifold; odd $N_f$ instead realizes the twofold Ising-odd
sector (Sec.~\ref{sec:baseline}). After flat-band
projection we build the single-band model with a fast momentum-resolved contraction of the
three-body matrix elements. The explicit Bloch matrix and the projected three-body matrix elements
are collected in Appendix~\ref{app:explicit}.

\subsection{Nonreciprocal skin term}

We add uniform nonreciprocal hopping along $a_1$, making the model non-Hermitian, with right- and
left-moving amplitudes differing by $e^{\pm\gamma}$ [Fig.~\ref{fig:overview}(a)]. In momentum space
this is the constant imaginary shift
\begin{equation}
  k_1 \;\longrightarrow\; k_1 + i\gamma ,
  \label{eq:shift}
\end{equation}
the Hatano--Nelson generator of the non-Hermitian skin effect~\cite{hatano1996,yao2018,okuma2020}.
The parameter $\gamma\ge0$ tunes away from the Hermitian point. The shift preserves lattice
translation, so total momentum remains a good quantum number and the many-body Hamiltonian remains
block diagonal~\cite{haldane1985,bernevig2012emergent}. On the torus the spectrum is complex and the
states remain extended; with an open $a_1$ boundary the right and left states accumulate at opposite
edges. Winding, open-boundary piling, similarity diagnostics, and the flat-band caveat are collected in
Supplemental Material Sec.~S5.

At each $\gamma$ we recompute the biorthogonal lowest-band eigenvectors of
$h(k_1+i\gamma,k_2)$ and project Eq.~\eqref{eq:Hint} onto their form factors
(Appendix~\ref{app:explicit}). The band-truncation ladder and solver checks are given in Supplemental
Material Sec.~S3.

\subsection{Biorthogonal six-state subspace and the non-Abelian counting}

For a non-Hermitian $H$ we solve the right and left eigenproblems
$H\rket{\psi_n}=E_n\rket{\psi_n}$ and $H^\dagger\lket{\psi_n}=E_n^{*}\lket{\psi_n}$, with
$\braket{\psi_m^{\mathrm{L}}}{\psi_n^{\mathrm{R}}}=\delta_{mn}$. Following non-Hermitian fractional
quantum Hall studies~\cite{yoshida2019}, the primary low-energy selection is the six
lowest-$\mathrm{Re}\,E$ eigenpairs. At $\gamma=0$ these form the Moore--Read ground-state manifold. Across
each counting window these instantaneous lowest-six eigenpairs coincide with the six-state branch continued from that point as $\gamma$ varies and remain separated by a real
line gap~\cite{gong2018topological,kawabata2019}. We form the equally weighted biorthogonal density
operator
\begin{equation}
  \rdm \;=\; \frac{1}{6}\sum_{n=1}^{6}\, \rket{\psi_n}\!\lbra{\psi_n} ,
  \label{eq:rho}
\end{equation}
which reduces to the standard $\frac{1}{6}\sum_n\ketbra{\psi_n}{\psi_n}$ at $\gamma=0$. This normalized
density operator is proportional to the rank-six spectral projector
($\rdm^2=\rdm/6$) and is invariant under a basis change within the selected subspace.

Its partial trace $\rho_A=\operatorname{Tr}_{\bar A}\rdm$ preserves the left/right expectation rule.
As consistency checks, we also reduce the right-state operator
$\rho_{\mathrm{RR}}=\frac{1}{6}\sum_n\rket{\psi_n}\rket{\psi_n}^{\dagger}$ and the normalized
right-subspace projector $Q_{\mathrm R}/6$, where
$Q_{\mathrm R}=R(R^\dagger R)^{-1}R^\dagger$. The former depends on the chosen right-state basis,
whereas the latter projects orthogonally onto its span and is basis invariant. The comparisons and the
open-boundary similarity check are given in Supplemental Material Secs.~S2 and S5.

We obtain the particle-entanglement spectrum by tracing out $N_f-N_A$ particles and writing the
reduced-operator eigenvalues $\lambda$ as $\xi=-\ln\lambda$~\cite{li2008,sterdyniak2011}. For
the biorthogonal density matrix these eigenvalues are generally complex, giving the non-Hermitian
entanglement spectrum~\cite{herviou2019,chang2020nh}. We use $N_A=4$ as the primary cut and $N_A=6$
as a balanced cross-check.

The Moore--Read fingerprint is the \emph{counting}: the number of PES levels below the
entanglement gap must equal the number of $(2,4)$-admissible configurations. These configurations place
$N_A$ particles on $N$ orbitals with at most $2$ particles in any $4$ consecutive orbitals
(cyclic on the torus)~\cite{read1999beyond,bernevig2008model,sterdyniak2011}. Enumerating this rule gives the
clean references
\begin{equation}
  \nbel^{\mathrm{ref}}(N_A{=}4) = 1308,\,3965,\,9282
  \quad (N=16,20,24),
  \label{eq:ref}
\end{equation}
with $\nbel^{\mathrm{ref}}(N_A{=}6)=70408$ at $N=24$. The same admissibility rule is
used by an independent Hermitian Moore--Read study~\cite{nonabelian_moire2024}, whose
$20$-orbital cluster gives $3965$, exactly matching our $N=20$ reference. Enumerating the
identical rule at $N=16$ and $N=24$ gives $1308$ and $9282$, which our exact diagonalization
reproduces at $\gamma=0$ (Sec.~\ref{sec:baseline}).

Our $N_A=4$ PES protocol uses two independent tests. First, the theory fixes
$n_{\mathrm{ref}}=\nbel^{\mathrm{ref}}$ before the spectrum is inspected, and we evaluate the gap
\begin{equation}
  \Delta_\xi^{\mathrm{ref}}
  =\mathrm{Re}\,\xi_{n_{\mathrm{ref}}+1}-\mathrm{Re}\,\xi_{n_{\mathrm{ref}}}
  \label{eq:refgap}
\end{equation}
at that predicted rank; the levels are ordered by $\mathrm{Re}\,\xi$, so every gap quoted here is a
difference of real ordered quantities. This tests whether the predicted low-lying subspace is separated; it does not
infer a count from a data-selected gap. Second, without supplying $n_{\mathrm{ref}}$, we scan every
consecutive gap in the full spectrum allowed by the stated convention and record the rank below the
widest one. Under the main convention, this blind rank equals $n_{\mathrm{ref}}$ at every sampled point
in each survival window.
The reference-rank gap and the blindly selected gap location therefore test the predicted rank in two
logically distinct ways. Their separation factors are reported in Supplemental Material Sec.~S2.

For reproducibility, the main biorthogonal convention at $\gamma>0$ keeps
$\mathrm{Re}\,\lambda>\epsilon$, with $\epsilon=10^{-12}$, and orders the retained levels by
$\mathrm{Re}(-\ln\lambda)=-\ln|\lambda|$. We also test the full $|\lambda|$ spectrum without a
discard, positive support, $-\ln\mathrm{Re}\,\lambda$, singular values, and the Hermitianized
operator. Within every reported window the reference count is unchanged by the tested reading or
reduced-operator construction. The eigenvalue census, comparison spectra, cutoff sweep, and
complex-level marginality are reported in Supplemental Material Sec.~S2.

\subsection{Energy isolation and competing states}

The global real-line gap $\Eg$ is the difference in $\mathrm{Re}\,E$ between the seventh eigenvalue and
the highest member of the selected six-state subspace. At $N=24$ it is computed over all $24$ momentum
sectors. Because a sixfold degeneracy alone does not exclude charge order, we also test the ordering
momenta, static structure factor, and orbital occupations~\cite{nonabelian_moire2024}; the results are
summarized in Sec.~\ref{sec:baseline} and Supplemental Material Sec.~S7.

\subsection{Solver}

The complex spectrum is obtained by exact diagonalization in each momentum sector, with dense solvers at
$N=16,20$ and a warm-started sparse Krylov solver at $N=24$. Convergence, band truncation, and
the all-sector protocol at the delocking points are documented in Supplemental Material Sec.~S3.

\section{Clean Moore--Read baseline}
\label{sec:baseline}

\begin{figure*}[t]
  \centering
  \includegraphics[width=\linewidth]{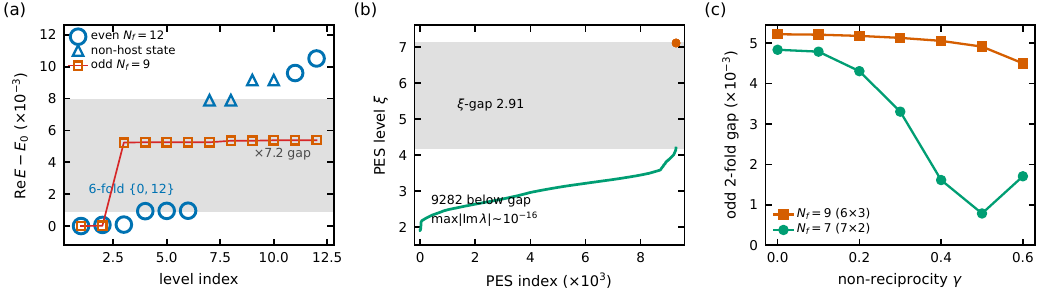}
  \caption{Clean Moore--Read baseline at $\gamma=0$ and its Ising even-odd fingerprint. (a) Low-lying
    many-body spectrum $E-E_0$ versus level index. Even fermion number ($N_f=12$, $N=24$; blue) gives the
    sixfold near-degenerate Moore--Read manifold in momentum sectors $\{0,12\}$ (circles below the shaded
    gap; gap-to-manifold-width ratio $7.2$); above the gap sit the first intruders from other momentum
    sectors (triangles) and higher host-sector states (circles). Odd fermion number ($N_f=9$, $N=18$;
    orange squares) instead gives a twofold Ising-odd manifold,
    the $d_\sigma=\sqrt2$ even-odd effect of the underlying $p$-wave pairing. (b)
    Particle-entanglement spectrum ($N_A=4$): the levels below the entanglement gap number exactly
    $9282$, equal to the $(2,4)$-admissible reference of Eq.~\eqref{eq:ref}; the gap is clean ($\xi$-gap $2.91$)
    and $\max|\mathrm{Im}\,\lambda|\sim10^{-16}$. (c) The odd twofold gap (to the third state) versus
    $\gamma$ for both odd sizes ($N_f=7$ on $7\times2$, $N_f=9$ on $6\times3$). The tracked doublet
    remains paired and separated through $\gamma=0.6$; at $N_f=7$ it is temporarily overtaken in the
    global $\mathrm{Re}\,E$ ordering near $\gamma=0.5$.}
  \label{fig:baseline}
\end{figure*}

Before adding non-Hermiticity we establish that the Hermitian model is a clean Moore--Read
state at all three sizes. At $\gamma=0$ the kagome flat band with the pure three-body
interaction of Eq.~\eqref{eq:Hint} has a sixfold near-degenerate ground manifold; at $N=24$ the
six states sit in momentum sectors $\{0,12\}$ with a gap-to-manifold-width ratio of $7.2$
[Fig.~\ref{fig:baseline}(a)]. The ground-state degeneracy is $6$ at each size, as required for
Moore--Read order on the torus.

This degeneracy carries a sharper non-Abelian signature in its dependence on the electron-number
parity. The Moore--Read state hosts an Ising anyon ($d_\sigma=\sqrt2$). Its torus
ground-state degeneracy is $6$ at even fermion number and $2$ at odd, the even-odd effect of the
underlying $p$-wave pairing~\cite{readgreen2000}; recent fractional-Chern-insulator studies use this
effect as a non-Abelian diagnostic~\cite{reddy2024,chen2024evenodd}. Our own $(2,4)$-admissibility
rule reproduces the split at the root level ($6$ for even $N_f$, $2$ for odd), and exact
diagonalization of the interacting model confirms it. At odd filling the ground manifold is a clean,
well-separated twofold: $N_f=7$ on a $7\times2$ torus, with the gap to the third
state $36$ times the twofold splitting; for $N_f=9$ on $6\times3$, the gap is $263$ times the splitting
[Fig.~\ref{fig:baseline}(a)]. This is an energy-level fingerprint of the non-Abelian Ising sector,
independent of the entanglement counting and not reproduced by a charge-density wave. It persists
under the non-Hermitian deformation at both odd sizes [Fig.~\ref{fig:baseline}(c)]. The adiabatically
tracked Ising-odd doublet remains internally paired and separated through the tested range
$\gamma\le0.6$. At $N_f=7$ it is temporarily overtaken by another doublet in the global
$\mathrm{Re}\,E$ ordering near $\gamma=0.5$; no such exchange occurs at $N_f=9$ (Supplemental
Material Sec.~S8).

The counting is the central diagnostic. The $N_A=4$ particle-entanglement spectrum has exactly
$1308$, $3965$, and $9282$ levels below the entanglement gap at $N=16,20,24$, respectively, matching the
$(2,4)$-admissible references of Eq.~\eqref{eq:ref} index by index
[Fig.~\ref{fig:baseline}(b)]. The provenance and independent-study comparison of these reference
counts are given with Eq.~\eqref{eq:ref} in Sec.~\ref{sec:model}. The
balanced $N_A=6$ cut gives $70408$, again equal to its reference. Resolved by the momentum $K_A$ of the
cut particles, the counting matches the $(2,4)$-admissible reference sector by sector through the
window edges $\gamma=0.55$ and $0.65$ at $N=16$ and $20$, respectively, and through
$\gamma=0.70$ at $N=24$. At the $N=24$ edge
$\gamma=0.74$, the total count $9282$ is verified (Supplemental Material Sec.~S1).
At $\gamma=0$ the entanglement spectrum is real to $\max|\mathrm{Im}\,\lambda|\sim10^{-16}$, as
it must be.

At $N=24$ the occupation remains near $\langle n_k\rangle=0.50$ throughout the window, while the Bragg ratio
$S(M)/\langle S(q{\neq}M)\rangle\approx0.75$--$0.79$ remains below unity. The full structure-factor and
occupation diagnostics, including the displaced figure, are in Supplemental Material Sec.~S7.

A $1$--$2\%$ two-body admixture does not remove the finite window at the tested sizes (Supplemental
Material Sec.~S7).

\section{Survival of the non-Abelian counting}
\label{sec:survival}

\begin{table*}[t]
  \centering
  \caption{Moore--Read $(2,4)$ counting survives the listed windows. Columns give orbital count $N$, torus
    $L_1\times L_2$, fermion count $N_f$, and momentum sectors of the sixfold subspace. Each window is the largest
    connected sampled interval from $\gamma=0$ over which every tested spectral reading and
    reduced-density-operator construction returns the reference count, with an open global
    momentum-sector-resolved gap. Adjacent-sample brackets mark PES delocking under the main
    convention ($\gpes$) and entry of a non-host sector into the six lowest-$\mathrm{Re}\,E$ states ($\gdem$).
    Last column: $\Delta_\xi^{\mathrm{ref}}$ at the window edge $\to$ first post-$\gpes$ sample.}
  \label{tab:sizes}
  \begin{ruledtabular}
  \begin{tabular}{lcccccccc}
   $N$ & $L_1\times L_2$ & $N_f$ & six-state sectors & $\nbel^{\mathrm{ref}}$ & survival window & $\gpes$ & $\gdem$ & $\Delta_\xi^{\mathrm{ref}}$: edge $\to$ post-$\gpes$ \\
   \hline
   $16$ & $4\times4$ & $8$  & $\{0\}$        & $1308$ & $\gamma\le0.55$ & $(0.60,0.65]$ & $(0.60,0.65]$ & $3.37\to1.74\times10^{-2}$ \\
   $20$ & $5\times4$ & $10$ & $\{0,5,10,15\}$ & $3965$ & $\gamma\le0.65$ & $(0.65,0.70]$ & $(0.65,0.70]$ & $1.60\to2.13\times10^{-12}$ \\
   $24$ & $6\times4$ & $12$ & $\{0,12\}$      & $9282$ & $\gamma\le0.74$ & $(0.74,0.75]$ & $(0.76,0.77]$ & $2.69\to1.90\times10^{-3}$ \\
  \end{tabular}
  \end{ruledtabular}
\end{table*}

\begin{figure}[!t]
  \centering
  \includegraphics[width=\linewidth]{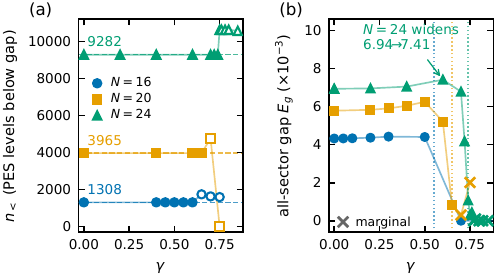}
  \caption{Survival of the Moore--Read $(2,4)$ counting under the skin effect.
    (a)~Number of PES levels below the entanglement gap, $\nbel$, versus nonreciprocity $\gamma$ for
    $N=16,20,24$ ($N_A=4$). The count remains at the clean references $1308/3965/9282$ (dashed lines)
    throughout the reported windows $\gamma\le0.55/0.65/0.74$ and changes at the first sampled point
    beyond the size-dependent $\gpes$. At $N=16$, the main convention still returns $1308$ at
    $\gamma=0.60$, outside the reported window, while sign-blind readings give a different result;
    $\gpes$ therefore remains $(0.60,0.65]$. Filled/open symbols denote samples locked/delocked under
    the main convention (Sec.~\ref{sec:coda}).
    (b)~Many-body real-line gap $\Eg$ from the six selected states to the
    seventh state. It remains open throughout each survival window; at $N=24$ it is evaluated over all
    $24$ momentum sectors. Vertical dotted lines mark the window edges, and crosses mark the delocked
    samples, where the complex-PES marginality $\mathcal{M}$ is no longer small (Supplemental Material,
    Sec.~S2).}
  \label{fig:survival}
\end{figure}

Moore--Read PES counting survives over a finite window. We define $\gpes$ as the
adjacent-sample bracket in which the PES of the instantaneous six lowest-$\mathrm{Re}\,E$ states
delocks from the fixed Moore--Read reference rank. Figure~\ref{fig:survival}(a) shows
$\nbel(\gamma)$ at the three sizes. The count remains at the clean reference of Eq.~\eqref{eq:ref}:
$1308$ for $\gamma\le0.55$ at $N=16$, $3965$ for $\gamma\le0.65$ at $N=20$, and $9282$ for
$\gamma\le0.74$ at $N=24$. At $N=16$, all $15$ combinations of three reduced operators and five
spectral readings return $1308$ at $\gamma=0$, $0.50$, and $0.55$. The next sample,
$\gamma=0.60$, lies outside this reading-independent window: the main convention still returns
$1308$, but sign-blind readings do not. Thus the window ends at $0.55$ while the main-convention
delocking bracket remains $\gpes\in(0.60,0.65]$. At the other window edges, the no-discard,
$|\lambda|$-ordered, and $\mathrm{Re}\,\lambda$-ordered readings all return $3965$ at $N=20$,
$\gamma=0.65$, and all five biorthogonal readings return $9282$ at $N=24$, $\gamma=0.74$
(Supplemental Material Sec.~S2).
Within each window the instantaneous-lowest-six states form one manifold continued smoothly in
$\gamma$, rather than unrelated states with the same count. Its complex energies
change by at most $3.1\times10^{-4}$ between adjacent samples, and consecutive spectral projectors have
near-unit overlap (Supplemental Material Sec.~S2). Table~\ref{tab:sizes} summarizes the results.

The selected states must also remain isolated. Figure~\ref{fig:survival}(b)
shows the real-line gap $\Eg$ to the seventh state. At $N=24$ we evaluate it over
\emph{all} $24$ momentum sectors, not a preselected set. The global gap rises
from $6.94\times10^{-3}$ at $\gamma=0$ to a peak $7.41\times10^{-3}$ at $\gamma=0.6$, and scanning
every sector rules out the host-only artifact that a momentum-incomplete sector set would produce
near a transition. Through the window edge $\gamma=0.74$ the lowest-six cluster still lies entirely in the host
sectors $\{0,12\}$. The lowest non-host sector enters the lowest six only past the window
(Sec.~\ref{sec:coda-b}); the global momentum-sector-resolved gap stays open across the
window (Supplemental Material Sec.~S3 and Table~S5).

The gap at the theoretically fixed reference rank stays open across every window
($\Delta_\xi^{\mathrm{ref}}\ge1.6$, Table~\ref{tab:sizes}), with the selected states remaining in their expected host sectors throughout each
window. The reference-rank gap is sharply reduced at the first samples beyond $\gpes$. The energy-level isolation of the six-state subspace, measured by the gap $\Eg$ to the seventh
state relative to its internal spread $\delta_6$, is a distinct, more marginal quantity: the ratio
$\Eg/\delta_6$ is $3.5/7.1/7.2$ for $N=16/20/24$ at $\gamma=0$, a property of the three-body interaction at these finite sizes, and it narrows with
$\gamma$. At the $N=20$ and $N=24$ window edges, $\Eg/\delta_6\simeq0.7$ and $0.8$; at $N=16$ it
reaches $0.3$ at the outside-window sample $\gamma=0.60$. The $N=24$ all-sector data are in
Supplemental Material Table~S5. The subspaces are well isolated ($\Eg/\delta_6\ge3$) through the
bulk of each window, and the reference-rank entanglement gap remains open at the edges.

The reference-index entanglement gap [Fig.~\ref{fig:coda}(a)] is more sensitive and depends on size. At $N=16$ it
rises across its window ($3.01\to3.37$, $+12\%$ at the edge $\gamma=0.55$). At $N=20$ and $N=24$ it changes
by $-43\%$ at the edge $\gamma=0.65$ and $-8\%$ at the edge $\gamma=0.74$, respectively (at a common
$\gamma=0.6$, the respective changes are milder: $-9\%$ and $-4\%$). At $N=24$ the change is
non-monotonic, dipping to $-15\%$ mid-window. The gap stays strictly positive, so the counting never delocks. We therefore use the count-lock, not the entanglement-gap trend, as the primary
diagnostic.

The finite counting window is not tied to the lattice direction $a_1$.
At $N=20$ the $5\times4$ torus breaks the $x\leftrightarrow y$ symmetry, so deforming along $a_2$
is an independent test. The count remains $3965$ through $\gamma\le0.65$ and delocks at
$\gamma=0.7$, where the same non-host sectors enter the lowest six (Supplemental Material Sec.~S7).

Geometry matters even at fixed size: for $N=24$ the $8\times3$ torus delocks in
$\gpes\in(0.62,0.64]$, below the $6\times4$ bracket $(0.74,0.75]$. This establishes geometry
dependence; the sector, gap, aspect-ratio, and circumference data are in Supplemental Material Sec.~S7.

\section{Destabilization beyond \texorpdfstring{$\gpes$}{gamma PES}}
\label{sec:coda}

\subsection{Counting destabilization}
\label{sec:coda-b}

\begin{figure}[t]
  \centering
  \includegraphics[width=\linewidth]{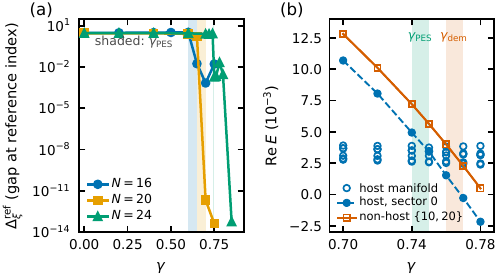}
  \caption{(a)~Instantaneous-lowest-six reference-rank entanglement gap
    $\Delta_\xi^{\mathrm{ref}}$ versus $\gamma$ for
    $N=16,20,24$ (log scale); shading marks each $\gpes$ bracket. (b)~$\mathrm{Re}\,E$ at $N=24$ for
    six plotted host-sector branches (open circles), a descending host-sector-$0$ branch (filled circles), and
    the lowest non-host pair $\{10,20\}$ (squares). Shaded intervals mark PES delocking,
    $\gpes\in(0.74,0.75]$, and non-host entry, $\gdem\in(0.76,0.77]$; instantaneous-lowest-six
    delocking precedes non-host entry.}
  \label{fig:coda}
\end{figure}

Beyond $\gpes$ the instantaneous-lowest-six count no longer matches the reference. Spectral readings
need not agree outside the reported windows, so there is no unique post-threshold integer. At $N=16$ and $\gamma=0.65$,
biorthogonal readings give ranks near $1700$, including $1741$, whereas the blind widest gap of
$Q_{\mathrm R}/6$ lies at rank $1452$. We therefore diagnose delocking through the gap at the
theoretically fixed reference rank. The rank is independent of the spectral reading, although the gap
value is not. Figure~\ref{fig:coda}(a) shows the main-convention gap remaining $O(1)$ throughout the
survival windows and becoming small at the first delocked samples. From the window edge to that sample
it falls by factors of $194$, $7.5\times10^{11}$, and $1400$ for $N=16$, $20$, and $24$. These are
sampled endpoint reduction factors. The $N=20$ value reaches the numerical floor
(Table~\ref{tab:sizes}). The $N=24$ global all-sector check is in Supplemental Material Sec.~S3.

Two thresholds must be distinguished. At $\gpes$, the reference-rank gap of the instantaneous
lowest-six subspace collapses and $\nbel$ leaves the reference. At $\gdem$, a non-host momentum sector
enters the six lowest-$\mathrm{Re}\,E$ states.
Each threshold is quoted as the interval between the last sample before and the first sample after the
event. The intervals coincide within the sampling resolution at $N=16,20$. At $N=20$, sectors
$\{6,19\}$ enter the lowest six at $\gamma=0.7$, the same sampled point at which the PES count
delocks. At $N=24$, an all-sector scan gives $\gpes\in(0.74,0.75]$ and
$\gdem\in(0.76,0.77]$. At the intervening samples $\gamma=0.75$ and $0.76$, the
instantaneous-lowest-six count is already delocked but all six lowest-$\mathrm{Re}\,E$ states remain
in host sectors $\{0,12\}$; the non-host
pair $\{10,20\}$ remains above them [Fig.~\ref{fig:coda}(b); Supplemental Material Sec.~S3 and
Table~S5]. Thus the $N=24$ instantaneous-lowest-six PES fingerprint delocks before a non-host momentum
sector enters the lowest six.

The low-lying spectrum nevertheless changes between the thresholds. A state in host sector $0$ enters
the lowest six in the same bracket $(0.74,0.75]$ as $\gpes$ and becomes the host-sector minimum from
$\gamma=0.76$ onward [Fig.~\ref{fig:coda}(b); Supplemental Material Table~S5]. Consequently, the
instantaneous six-state sets at $\gamma=0.75$ and $0.76$ are not uniquely identified with the six states
continued from $\gamma=0$.

Within the $\gpes$ bracket a sector-$0$ branch pair leaves the real axis. On the sampled grid
($\Delta\gamma=10^{-3}$) the two levels are real and split by $2.9\times10^{-4}$ at $\gamma=0.744$;
from $\gamma=0.745$ to $0.749$ their real parts agree to $10^{-16}$ while their imaginary parts are
equal and opposite, reaching $\pm1.94\times10^{-4}$ at $\gamma=0.747$; by $\gamma=0.750$ they are two
separated real levels again. Across the interval the branch-tracking
overlaps become nearly degenerate---the best and second-best matches differ by a factor $0.95$ at
$\gamma=0.745$---and the continued-branch identity is therefore two-valued beyond the interval. At $\gamma=0.75$ the two partners give different PES results: the main
continuation delocks with count $10616$, whereas the alternate continuation locks to $9282$ with
$\Delta_\xi^{\mathrm{ref}}=3.012$, continuing the in-window trend. At the sampled points
$\gamma=0.76,0.77,0.80$, both continuations delock. Thus the original-subspace verdict is ambiguous at
$0.75$ but branch-choice independent at these later samples; Supplemental Material Sec.~S2 compares both
subspaces and gives the tracking audit.

The balanced $N_A=6$ cut provides a secondary check that is insensitive to the tail gap. It uses the fixed reference rank $70408$, and its
reference-index entanglement gap stays open through the survival window and closes to
$1.7\times10^{-14}$ at $\gamma=0.8$. At $\gamma=0.4$ the count below the single largest gap migrates to a
tail singleton ($134595$), but the reference-index count remains $70408$ (Supplemental Material
Sec.~S2).

Supplemental Material Sec.~S2 gives the complex-PES marginality and reading audit, including the
negative-real-part tail and reading dependence at the outside-window sample $N=16$, $\gamma=0.60$.

\subsection{Reordering of \texorpdfstring{$\mathrm{Re}\,E$}{Re E} between momentum sectors}
\label{sec:coda-ep}

At $\gdem$, states from non-host sectors enter the six lowest-$\mathrm{Re}\,E$ levels:
$\{1,2,3\}$, $\{6,19\}$, and $\{10,20\}$ for $N=16,20,24$, respectively. The skin term preserves
translation, so momentum remains a good quantum number and the many-body Hamiltonian is block diagonal.
States in distinct momentum blocks have no matrix element between them and do not hybridize.

On the sampled grid, the later host--non-host event at $\gdem$ changes only the ordering of the real
parts. At $N=24$ the plotted host-sector levels satisfy $|\mathrm{Im}\,E|\le2.3\times10^{-4}$ through $\gamma=0.78$, whereas the descending
pair $\{10,20\}$ has $\mathrm{Im}\,E=\mp6.4\times10^{-3}$ near the reordering. The smallest sampled
complex-plane separation between the host-sector levels and this pair is $6.3\times10^{-3}$ at
$\gamma=0.76$, more than
three times the spread of the six lowest levels there (Supplemental Material Table~S5 and Fig.~S4).
No sampled approach to a complex-energy degeneracy accompanies this host--non-host reordering. This
statement does not include the same-sector complex-conjugate interval at $\gpes$ described above. The lowest-six
selection nevertheless changes at $\gdem$ because it is defined by $\mathrm{Re}\,E$.

\subsection{Abelian versus non-Abelian in the same lattice}
\label{sec:coda-abelian}

We compare with a $\nu=\third$ Laughlin state in the same kagome lattice, nonreciprocal deformation,
and code. For $N=24$, $N_f=8$, and a two-body interaction, its $(1,3)$ count remains $2730$ through
$\gamma=1.0$, whereas the Moore--Read lowest-six count delocks in $(0.74,0.75]$. This is a model-level,
multi-parameter comparison: it changes the filling, fermion number, and interaction order, and its
$\gamma=0$ real-line gap is $6.7\times10^{-2}$ rather than the Moore--Read value
$6.94\times10^{-3}$. It therefore shows only that this particular Laughlin realization is the more robust
under the same dimensionless deformation, not that Abelian order is generically more stable than
non-Abelian order. Supplemental Material Sec.~S4 gives the gap, spectral-reading, and literature comparison.

\section{Conclusion and outlook}
\label{sec:conclusion}

We have studied a non-Hermitian Moore--Read fractional Chern insulator using the instantaneous
lowest-six subspace as the primary PES diagnostic under a skin-generating deformation. A finite-size
survival window occurs: across
three sizes the Moore--Read $(2,4)$ entanglement counting locks to the clean references
$1308/3965/9282$ over a finite range of nonreciprocity, and the defining entanglement gap stays finite
and positive across each window. Beyond the geometry-dependent delocking threshold
$\gpes$, the instantaneous-lowest-six counting destabilizes and its reference-rank gap drops sharply
across the sampled bracket. At $N=24$ the delocking at
$\gpes\in(0.74,0.75]$ precedes the exchange of $\mathrm{Re}\,E$ ordering between momentum sectors at
$\gdem\in(0.76,0.77]$; within the $\gpes$ bracket a sector-$0$ branch pair becomes complex conjugate, after
which the $\gamma=0.75$ continuation is two-valued, while both continuations delock at the tested PES points
$\gamma=0.76$, $0.77$, and $0.80$. The later host--non-host families remain separated in complex
energy at the sampled points (Sec.~\ref{sec:coda}). In the same lattice, a
$\nu=\third$ Laughlin realization keeps its counting locked to $\gamma=1.0$, so its counting is the
more robust (Sec.~\ref{sec:coda-abelian}).

Beyond this specific model, the biorthogonal Moore--Read counting is a diagnostic for the
fate of a non-Abelian manifold under any momentum-preserving non-Hermitian deformation. To our
knowledge, this is its first application. The non-Abelian diagnosis combines the $(2,4)$ counting with
the tracked Ising even--odd pattern over the tested range $\gamma\le0.6$.

Several extensions remain open. A biorthogonal infinite-cylinder iDMRG calculation could test modular
data and quasihole statistics beyond the PES~\cite{zaletel2013,cincio2013}. The survival threshold depends on geometry as well as size: at fixed $N=24$ the $8\times3$ torus,
with larger aspect ratio and narrower transverse circumference, delocks earlier than the $6\times4$
one (Sec.~\ref{sec:survival}), while along the only sampled size sequence, the fixed-$L_2$ tori
$4\times4\to5\times4\to6\times4$, the reading-independent window endpoints increase from
$0.55$ to $0.65$ to $0.74$. The distinct main-convention brackets move from
$\gpes\in(0.60,0.65]$ to $(0.74,0.75]$. No sampled comparison shows the window shrinking as the system
grows, but in both comparisons the two torus dimensions change together, so fixed-aspect
thermodynamic scaling remains undetermined. Resolving it still requires a fixed-aspect even-$N_f$
sequence at sizes beyond the reach of exact diagonalization.
If the reading-independent window survives fixed-aspect scaling, its endpoint would quantify a finite
tolerance to nonreciprocity, while $\gpes$ would locate delocking under the main convention.
Dissipative Laughlin and cold-atom Pfaffian results supply the non-Hermitian and non-Abelian ingredients,
respectively, in separate settings~\cite{yoshida2019,pfaffian_bosons2026}. A matched Abelian control would sharpen the
comparison, and we expect the survival-then-reorganization picture and the biorthogonal counting
diagnostic to carry over to the broader Read--Rezayi series and to other nonreciprocal deformations.

\begin{acknowledgments}
We gratefully acknowledge support from the Volkswagen Group of America. F.~B.~P. received support through a contract with the Volkswagen Group of America. J.~G. was supported by the Walecka Fellowship, S.~N. by the Stanford NPL Fund, and X.~Z. by the Timothy Francis Kennedy Memorial Scholarship.
\end{acknowledgments}

\section*{Data availability}
The data that support the findings of this article are available from the authors upon
reasonable request.

\appendix
\renewcommand{\thefigure}{A\arabic{figure}}
\setcounter{figure}{0}
\section{Explicit Bloch Hamiltonian and projected interaction}
\label{app:explicit}

\emph{Bloch Hamiltonian.} With the three kagome sublattices labeled $A,B,C$, the tight-binding
Hamiltonian is the Hermitian $3\times3$ Bloch matrix $H_0(k)$ with zero diagonal and upper-triangular
entries
\begin{equation}
\begin{aligned}
\left[H_0(k)\right]_{AB} &= -(t_1-i\lambda)\,\bigl(1+e^{-ik_x}\bigr),\\
\left[H_0(k)\right]_{AC} &= -(t_1+i\lambda)\,\bigl(1+e^{-ik_y}\bigr),\\
\left[H_0(k)\right]_{BC} &= -(t_1-i\lambda)\,\bigl(1+e^{i(k_x-k_y)}\bigr),
\end{aligned}
\label{eq:bloch}
\end{equation}
and $\left[H_0\right]_{\beta\alpha}=\left[H_0\right]_{\alpha\beta}^{*}$, with $k=(k_x,k_y)$ and
$t_1=\lambda=1$. The real part $t_1$ is the nearest-neighbor hopping and the imaginary part $\lambda$
is the chiral nearest-neighbor term that opens the topological gap~\cite{sun2011,tang2011}; the two
combine on each bond into $t_1\mp i\lambda$ as fixed by the bond orientation. Band flattening sets each
of the three bands to its $k=0$ energy while retaining the Bloch eigenvectors. The lowest band is thus exactly flat,
with all momentum dependence in its eigenvector, whose sublattice components are
$u_\alpha(k)=\langle\alpha|u_0(k)\rangle$. The nonreciprocal deformation is the component shift
$k_1\to k_1+i\gamma$ of Eq.~\eqref{eq:shift}, equivalently $k_x\to k_x+i\gamma$ in the coordinates used
here, under which $u_\alpha$ becomes
the biorthogonal pair $u^{\mathrm R}_\alpha,u^{\mathrm L}_\alpha$ of Sec.~\ref{sec:model}, normalized
$\braket{u^{\mathrm L}_k}{u^{\mathrm R}_k}=1$ with the reciprocal interaction evaluated at real lattice
momentum. Across the survival windows the lowest-band right eigenvector has minimum overlap $0.939647$
with its Hermitian counterpart on the discrete torus momenta and $0.9389$ on a dense mesh; the dense-mesh
minimum separation from the upper bands is $2.026919$, so no level reordering occurs.

\emph{Projected three-body interaction.} With $\creo{k}$ the lowest-band fermion creation operator at
momentum $k$, the interaction of Eq.~\eqref{eq:Hint} is
\begin{equation}
H_{\mathrm{int}} = U_3\!\!\sum_{\substack{k_1k_2k_3\\ k_4k_5k_6}}\!\!
\delta_G\, W\,
\creo{k_4}\creo{k_5}\creo{k_6}\,\anno{k_3}\anno{k_2}\anno{k_1},
\label{eq:h3}
\end{equation}
where $\delta_G$ enforces conservation of total momentum modulo a reciprocal-lattice vector. The
lowest-band-projected matrix element $W\equiv W(k_4k_5k_6;k_1k_2k_3)$ is the antisymmetrized product of
form factors
\begin{equation}
\begin{aligned}
W ={}& \frac{1}{N}\sum_{P,P'\in S_3}\mathrm{sgn}(P)\,\mathrm{sgn}(P')\\
&\times\, \bar u_A(k_{P'_4})\,\bar u_B(k_{P'_5})\,\bar u_C(k_{P'_6})\\
&\times\, u_C(k_{P_3})\,u_B(k_{P_2})\,u_A(k_{P_1})\,\Phi_{PP'},
\end{aligned}
\label{eq:w3}
\end{equation}
with $u\equiv u^{\mathrm R}$ and $\bar u\equiv(u^{\mathrm L})^{*}$ the biorthogonal lowest-band form
factors (reducing to the Hermitian Bloch eigenvector at $\gamma=0$), $P,P'$ the six permutations of the
annihilation and creation momenta, and
\begin{equation}
\Phi_{PP'} = 1+\exp\!\Bigl[-i\bigl((k_{P_2}-k_{P'_5})_x+(k_{P_3}-k_{P'_6})_y\bigr)\Bigr]
\label{eq:phi3}
\end{equation}
the smallest-triangle contact structure factor that selects the shortest-range relative
configuration. Equations~\eqref{eq:h3}--\eqref{eq:phi3} give a lattice analogue of the Moore--Read
three-body parent pseudopotential~\cite{moore1991,read1999beyond}; the two-body projector $P_2$ has the
same structure with two form factors and a single-bond structure factor. Throughout $U_3=1$ and
$U_2=0$.

The lowest band carrying the projected interaction is a $C=+1$ Chern band (Fukui--Hatsugai--Suzuki
at $\gamma=0$). Before flattening, its direct/indirect gaps are $2.27/1.27$ and flatness is $\Delta_{\mathrm{gap}}/W=1.0$.
Over the sampled survival-window range $\gamma\le0.74$, the discrete-torus and dense-mesh
overlap minima and the dense-mesh band-separation minimum are those summarized above; the
discrete-torus overlap minimum occurs at
$\gamma=0.725$, while the dense $96^2$ mesh, sampled in steps of $0.025$, places its overlap minimum at
$\gamma=0.70$.
Extending the dense-mesh isolation check to $\gamma=1.0$, the minimum separation falls to $1.700$ at
$\gamma=1.00$, with no closure over the interval. On the $6\times4$ momenta used for the Abelian
control, the minimum separation is $1.873$ and the minimum overlap with the Hermitian lowest-band right
eigenvector is $0.887$. The projected lowest band remains isolated through $\gamma=1.0$. A separate
exceptional point near $\gamma\approx0.59$ lies between the two discarded upper bands.

\bibliography{references}

\end{document}